\documentclass[onecolumn,aps,pre,nofootinbib]{revtex4-2}

\usepackage{amsmath}

\usepackage{fancyhdr}

\usepackage{physics}

\usepackage{natbib}

\begin{document}

\title{Assembling Cartesian Harmonic Tensors}
\author{William C. Parke}
\affiliation{Professor of Physics Emeritus \\The George Washington University, Washington, DC 20052}
\date{Feb 11, 2023}
\begin{abstract}
Cartesian harmonic tensors are completely symmetric traceless tensors in three dimensional space constructed from the direct product of a unit vector with itself.  They are
useful in generating expressions for the angular coupling of spherical harmonics of differing solid angles in terms of scalar and vector products of the unit vectors
for the spherical harmonics involved. A presentation for the algebraic programming of Cartesian harmonic tensors of arbitrary rank
is given using Algebraic Assembly Language. Results are checked using
Mathematica.


\end{abstract}

\maketitle

\fancyhf{}

\renewcommand{\footrulewidth}{0pt}

\renewcommand{\headrulewidth}{0pt}


\rfoot{{\tiny }}

\affiliation{Physics Department, The George Washington University,
Washington, DC 20052}

\section{Introduction}

The Cartesian spherical harmonics of this paper were described
by D.R. Lehman and the present author in 1989.\cite{lehman1989angular}
(Earlier considerations can be found in the work of
Coope et al.\cite{coope1965irreducibleI,coope1970irreducibleII,coope1970irreducibleIII}
Our paper adds closed-form expressions for the coefficients
in the decomposition of tensor products expansion, analogous
to those of Racah\,\cite{fano1959irreducible,racah1942theoryI,racah1942theoryII},
for products of Cartesian harmonic tensor couplings
and gives explicit expressions for the
coupling of up to five spherical harmonics to zero in terms of
dot products of the unit vectors of the harmonics.) At the time,
we were interested in calculating matrix elements involving
the expectation value of an operator for a three-particle
quantum system.  The wave functions for the three-particle system
can be expressed in terms of a sum over a product of two
spherical harmonics (one for the orientation of a pair of particles
and a second
for the angles of the third particle relative to the pair).
The matrix elements appear as $\bra{\psi_f}\mathcal{O}\ket{\psi_i}$.
The operator $\mathcal{O}$ may itself involve spherical harmonics.
We needed expressions for the coupling of up to five spherical
harmonics to produce a scalar, and we had hundreds
of such couplings. Of course, we could have
programmed the sums involved (e.g., in FORTRAN), and let the numerics take over.
However, there are cases in which large terms
of the expanded matrix element algebraically canceled,
but, with finite precision numerics, that cancellation would not
occur. We wanted an algebraic expression to do the sum
of the terms of the expanded matrix elements first, and
then do the numerics. By expressing the
polar spherical harmonic couplings in term of Cartesian spherical
harmonics, we gained the added benefit of having the
algebraic results of spherical harmonic coupling to a scalar in terms of the unit vectors in the spherical harmonics. We decided to
do the algebra in a program language called REDUCE begun in the 1963 by Tony Hern\,\cite{REDUCE1963,hern1983reduce} to simplify the trace of a product of Dirac gamma matrices, but evolved
into more general algebraic evaluations and simplifications.

Since the advent of electronic computers, a number of programs
characterized
as being `Computer Algebra Systems' (CAS) have matured. They act on symbols
and numbers and most are able to produce simplifications and evaluations
in symbolic form together with outputting rational numbers
as exact fractions. Prominent among these 
are the following:
\begin{center}
\begin{tabular}{lclll}
\hline
Program &\ Date\ \ \ & Originally written in\ \ \ \ \ &Similar to & Current Licensing  \\
\hline
FORTRAN  & 1957 & IBM 704 machine code\ \ & - & GNU version open \\
Lisp  & 1960 & IBM 704 machine code\ \ & - & GPL open source  \\
ALGOL 60 & 1960 &  Electrologica X1 code & - & GPL (ALGOL 68) \\
REDUCE  & 1963 & ALGOL 60, RLisp & - & BSD open source  \\
Axiom & 1965 & Lisp, FORTRAN, C\ \ \ & - & GPL open source  \\
APL  & 1966 & IBM 360 machine code  & - & GPL open source \\
Macsyma & 1968 & Lisp & - & commercial   \\
MATLAB & 1979 & C, FORTRAN & - & commercial  \\
Maple  & 1980 & Ada (Pascal), C & - & commercial  \\
Maxima  & 1982 & Lisp  & Maple & GPL open source  \\
Derive  & 1988 & $\mu$Lisp & Maple & Held by Texas Inst.\\
Mathematica  & 1988 & C, C++ & (unique) & commercial  \\
$\mu$PAD  & 1989 & C++ \ \  & Maxima & commercial  \\
Scilab & 1993 & C & MATLAB \ \ & CeCILL open \\
Scientific Workplace & 1993 & Pascal, Maple, $\mu$PAD \ \ & (Maple)* & commercial in UK  \\
Octave & 1994 & C++ & MATLAB\ & GPL open source  \\
NAR2000   & 1998 & C  & APL &  Free software \\
SageMath & 2005 & Python, Cython\ \ \ & MATLAB & GPL open source  \\
Reduce.jl  & 2012 & Reduce+Julia scripting\ \ & REDUCE & free MIT License \\
\hline
\multicolumn{5}{c}{* \rule[-1mm]{0mm}{7mm}Scientific Workplace v5.5 was made to edit LaTeX code but also contains a Maple engine.}\\
\end{tabular}
\end{center}

The purpose of getting results in algebraic form from a CAS program
includes to
\begin{itemize}
\setlength\itemsep{-0.2em}
\item simplify and algebraically evaluate initially complicated
expressions;
\item transfer algebraic results into other algebraic, numeric, or
graphical programs and text processors;
\item solve a system of equations expressed algebraically;
\item find closed-form expressions for doable sums and integrals;
\item gain insights about initially less scrutable expressions;
\item act on symbolic data streams
for decision-making and control systems.
\end{itemize}

The listed programs are sophisticated and extensively elaborated systems,
some with hundreds of developers and hundreds of thousands of users among engineers and scientists. Having the programs written in a higher level
language such as C means that developers can maintain and enhance
these programs with short turnover times, and be able to
quickly port the programs to a number of platforms.
However, there are important reasons
to be able to optimized CAS programs with
`algebraic assembly language' segments. In particular, these optimized
segments can replace bottlenecks and time-consuming parts.

As a demonstration of algebraic assembly language programming, the algebraic forms for Cartesian harmonic tensors
will be generated and then compared to
coding using Mathematica\,\cite{wolfram}.

\section{Applying Algebraic assembly language programming}

The Cartesian harmonic tensors can be formed from the direct products
of a unit vector (here called $\widehat{\boldsymbol{a}}$, which we
take to have polar angle $\theta$ and azimuthal angle $\phi$.). The direct-product tensor is then
made completely symmetric and traceless in all its tensor indices.
The resulting tensors of given rank becomes an irreducible representation of the rotation group $O(3)$.
These tensors correspond to the traditional spherical harmonics\,\cite{edmonds1957angular}
$Y^{\ell}_m(\theta,\phi)$, but with their base vectors
transformed from polar to Cartesian
form.

The components of the Cartesian harmonic tensors of rank $\ell$  constructed from the unit vector $\widehat{\boldsymbol{a}}$ are given by:
\begin{equation}
a^{\left\{ \ell\right\} }=\left[ \frac{\left( 2\ell-1\right) !!}{\ell!}\right]
\sum_{r=0}^{\lfloor \ell/2 \rfloor }\left( -1\right) ^{r}\left[ \frac{\left(
2\ell-2r-1\right) !!}{\left( 2\ell-1\right) !!}\right] \left\{ a\cdots \left(
\ell-2r\right) \cdots a\delta \cdots \left( r\right) \cdots \delta \right\} \ , \label{eq-aL}
\end{equation}
where explicit tensor indices on the $a$'s and on the Kronecker deltas
$\delta$ have been suppressed. The summation over $r$ goes from zero
to the first
integer at or just below $\ell/2$, denoted with `floor' brackets
$\lfloor\   \rfloor$.
The tensor $ \boldsymbol{a}^{\{\ell\}}$ is normalized to make its tensor contraction
with the vector $\widehat{\boldsymbol{a}}^{\{1\}}\equiv\widehat{\boldsymbol{a}}$ give the corresponding next lower rank tensor.
The pair of curly braces on the right-hand-side
indicate that the terms within are made completely symmetric
in their indices, with each distinct component represented
only once. The parenthetical number between the continuation
dots is the number of factors ($a$'s or $\delta$'s) present.

For example, the first four Cartesian harmonic tensors are
\begin{equation}
a^{\{1\}}_i=a_i  \ ,
\end{equation}
\begin{equation}
a^{\{2\}}_{i_1i_2}=\frac{3}{2}\left(a_{i_1}a_{i_2}
-\frac{1}{3}\delta_{i_1i_2}\right)  \ ,\label{eqa2}
\end{equation}
\begin{equation}
a_{i_{1}i_{2}i_{3}}^{\left\{ 3\right\}
}=\frac{5\cdot 3}{3\cdot 2}\left(a_{i_{1}}a_{i_{2}}a_{i_{3}}-\frac{1}{5}
\left(a_{i_{1}}\delta
_{i_{2}i_{3}}+a_{i_{2}}\delta _{i_{3}i_{1}}+a_{i_{3}}\delta
_{i_{1}i_{2}}\right)\right)  \ , \label{eqa3}
\end{equation}
\begin{eqnarray}
a_{i_{1}i_{2}i_{3}i_{4}}^{\left\{4\right\}
}&=&\frac{7\cdot 5\cdot 3}{4\cdot 3\cdot 2}\Big(a_{i_{1}}a_{i_{2}}a_{i_{3}}a_{i_4} \nonumber \\
&&\hspace{-1.2in}-\frac{1}{7} \left(a_{i_1}a_{i_2}\delta_{i_3i_4}+a_{i_1}a_{i_3}\delta_{i_2i_4}
+a_{i_1}a_{i_4}\delta_{i_2i_3}+a_{i_2}a_{i_3}\delta_{i_1i_4}+a_{i_2}a_{i_4}\delta_{i_1i_3}+a_{i_3}a_{i_4}\delta_{i_1i_2}\right) \nonumber \\
&&\hspace{-0.6in}+\frac{1}{7\cdot 5}
\left(\delta_{i_1i_2}\delta_{i_3i_4}+\delta_{i_1i_3}\delta_{i_2i_4}+\delta_{i_1i_4}\delta_{i_2i_3}\right)
\Big) \ . \label{eqa4}
\end{eqnarray}

The number of terms in any one pair of curly braces $\{\}$ will be
\begin{equation}
N_a(\ell,r)=\frac{\ell!}{(\ell-2r)!\,2^rr!}
=\frac{\Gamma(\ell+1)}{ \Gamma(\ell-2r+1)\,2^r\,\Gamma(r+1))} \ .
\label{eq-Na}
\end{equation}
Table \ref{tab-num-terms} shows how many terms there are in a given brace pair.
$N_a(10,r)$ reaches a maximum of $4899.4153802509$ at $r=3.7006759$.
The nearest $r$ integer is $4$, for which $N_a(10,4)=4,735$.
The number of terms is largest when $r$ is one below $\lfloor \ell/2\rfloor$, growing faster than exponentially:
$N_a\left(\ell,(\lfloor \ell/2 \rfloor -1)\,\right) \propto\exp{\ell\,\ln{(\ell)}}$).

\begin{center}
\begin{table}
\caption{Number of terms in a pair of braces}
\begin{tabular}{r r r r}
\hline
$\ell$ &   $r$  & terms in $\{\}$ & total terms in $a^{\{\ell\}}$\\
\hline
 6  &  2  &              45 &               75\\
 7  &  2  &             105 &              231\\
 9  &  3  &           1,260 &            2,629\\
10  &  4  &           4,735 &            9,495\\
13  &  5  &         270,270 &          568,503\\
14  &  6  &         945,945 &        2,390,479\\
19  &  8  &   1,964,187,225 &    4,809,701,439\\
20  &  8  &   9,820,936,125 &   23,758,664,095\\
21  &  9  &  45,831,035,250 &  119,952,692,895\\
22  &  10 & 151,242,416,325 &  618,884,638,911\\
\hline
\end{tabular}
\label{tab-num-terms}
\end{table}
\end{center}

However,
the number of independent components of the tensor $\boldsymbol{a}^{\{\ell\}}$
is limited because of symmetry and trace conditions, which
reduce the number to $(2\ell+1)$.
This can be seen as follows: For a symmetric tensor $\boldsymbol{T}^{\{\ell\}}$ of rank $\ell$ in a $d-$dimensional space,
spread $\ell$ dots in a row to represent the tensor indices of a given
component of $\boldsymbol{T}^{\{\ell\}}$, ordered
from smallest to largest. Values
of the indices may repeat. Put $d-1$ partitions on the left, between, or to the right of the dots and already placed partitions. Any dots to the left of the first partition will represent a value for
the tensor's first index, labeled $i_1$; any
between the first and second partition will represent a value for the second tensor index, labeled $i_2$, with $i_2\ge i_1$ etc. For example, if the rank of the tensor is four in a
space of six dimensions, then
$(\ \ |\ \bullet\ \bullet\ |\ \ |\ \bullet \ |\ \ |\ \bullet\ \ )$
would represent
the tensor components of the form $T^{\{4\}}_{i_2 i_2 i_4 i_6}$. The number of
ways to have placed the partitions is $(l+1)(l+2)\cdots (l+d-1)$. Since
the order in which the partitions are placed is immaterial, the
number of such distinct tensor components will be $(\ell+d-1)!/(\ell!\,(d-1)\,)!$,
which is the binomial coefficient $\binom{\ell+d-1}{\ell}$. The trace-%
vanishing conditions can be expressed as $\sum_k T_{kki_3i_4...}=0$.
There are $\binom{\ell-2+d-1}{\ell-2}$ such conditions, leaving
$\binom{\ell+d-1}{\ell}-\binom{\ell-2+d-1}{\ell-2}$ independent components. For $d=3$, this
becomes $(2\ell+1)$, the same as the number of $m$ values that
$Y^{\ell}_m(\theta,\phi)$ can have.

\section{Producing Cartesian Harmonic Tensors in Assembly Language}

For programming the generation of a Cartesian harmonic tensor, it is convenient to use the following
indexing scheme. Let `$\alpha$' be an integer from one to $\ell-2r$, used to number the indices on the `$a$' factors, with
$i_{\alpha}$ as the index of the `$\alpha$-th' `$a$' factor. Let `$\beta$' be an integer from one to $r$, used to number the indices on the $\delta$ factors, with $j_{\beta}$ as the first index on the
`$\beta$-th `$\delta$' factor,
and $k_{\beta}$ as the second index on the same `$\delta$'. The ranges of the indices of `$a$'
and the `$\delta$'
are given by
\begin{eqnarray}
  i_{\alpha-1} + 1 \le & i_{\alpha} & \le \ 2r+\alpha \ ,\\
  j_{\beta-1} + 1 \le & j_{\beta} & \le \ \ell-2r+2\beta -1\ ,\\
  j_{\beta}+1\le & k_{\beta} & \le \  \ell \ ,
\end{eqnarray}
where $i_0 \equiv 0$, $j_0 \equiv 0$ are taken as initial values
in the program. The first non-zero value for an $i$ or a $j$
index will be one and never greater than $\ell$.

Programming in assembly, of course,
is tedious, time consuming, and often obscure, which is why it is largely avoided, except in those cases requiring fast operations.  The advantage
of the assembly route is speed. At the machine level, one can
avoid unnecessary steps that are introduced by compilers in higher-level
languages. Those languages largely favor expediency over efficiency. The resultant speed advantage of assembled machine code can
sometimes be orders of magnitude over coding the same task
in a higher-level language. As a side benefit, while applying assembly coding
to a programming task, we can gain further intimate knowledge of the
inner workings and capabilities of a computer and the programs it
runs, as well as using knowledge of the CPU, associated
coprocessors, and device controllers, to make more effectual, faster and
tighter code.

In Appendix \ref{app-ass-code}, or in the ancillary
material associated with this paper, an assembly code (that runs
on a Ubuntu i386 system at the command-line level) is given to calculate the components of a Cartesian spherical harmonic tensor in algebraic form. Here are two example
of the code output:

\vspace{0.1in}

An output for $\ell=4$, (i.e. with the command ./cart 4) is

\noindent\hrulefill

\begin{verbatim}
a{4} :
+a1.a2.a3.a4
-(1/7)(a3.a4.d12 + a2.a4.d13 + a2.a3.d14 + a1.a4.d23 + a1.a3.d24 + a1.a2.d34)
+(1/(7.5))(d12.d34 + d13.d24 + d14.d23)

Number of terms in the tensor is 10
\end{verbatim}
\vspace{-0.1in}

\noindent\hrulefill

\vspace{0.2in}

An output for $\ell=5,\ r=2$ (i.e. with the command ./cart 5 2) is

\noindent\hrulefill

\begin{verbatim}
a{5,2} :
+(1/(9.7))(a5.d12.d34 + a4.d12.d35 + a3.d12.d45 + a5.d13.d24
 + a4.d13.d25 + a2.d13.d45 + a5.d14.d23 + a3.d14.d25 + a2.d14.d35
 + a4.d15.d23 + a3.d15.d24 + a2.d15.d34 + a1.d23.d45 + a1.d24.d35
 + a1.d25.d34)

Number of terms in the symmetry brace is 15
\end{verbatim}
\vspace{-0.1in}
\noindent\hrulefill

\vspace{0.1in}

For simplicity in form and beauty, the overall normalization factor $(2\ell-1)!!/\ell !$ is intentionally left out of the output.
Redirection (e.g. ./cart 5 $>>$\,cart-results.txt) will add the results to a text file. A text editor can then be used to put the results into
the format needed for subsequent programs, such as FORTRAN or REDUCE, that can act on them.

With REDUCE, the coupling of spherical harmonics can then
be found in algebraic form, such as\,\cite{lehman1989angular}:
\begin{equation} \begin{array}{l}
\left[ \left[ \left[ Y^{\left[ 1\right]} (a)\times Y^{\left[ 3\right]} (b)\right]
^{\left[ 2\right]} \times Y^{\left[ 2\right]} (e)\right] ^{\left[ 1\right]}
\times \left[ Y^{\left[ 1\right]} (d)\times Y^{\left[ 1\right]} (e)\right]
^{\left[ 1\right]} \right] ^{\left[ 0\right]} =3\sqrt{3}/(64\pi ^{5/2})\\
\{5(a\cdot b)(b\cdot c)(b\cdot d)(c\cdot e)-5(a\cdot b)(b\cdot c)(b\cdot e)(c\cdot d)\\
-(a\cdot c)(b\cdot d)(c\cdot e) +(a\cdot c)(b\cdot e)(c\cdot d)
-(a\cdot d)(b\cdot c)(c\cdot e) +(a\cdot e)(b\cdot c)(c\cdot d)\} \, . \label{A24}
\end{array} \end{equation}
(In this expression, we use the notation of Danos\,\cite{danos1972fully,danos1990angular}, who advocated using a phase
choice for $Y^{\left[\ell\right]}$ which makes couplings in quantum
matrix elements far simpler to handle.)

\section{Mathematica Code to Make Cartesian Harmonic Tensors}

As a check of the results of the assembly program, Mathematica code was used to generate the terms within the
symmetrization curly braces in the Cartesian harmonic tensor presented in Eq. (\ref{eq-aL}). The case $r=0$ is simplest, as the product of $\ell$ $a$'s (with different indices)
is already fully symmetric in any pair of those indices, i.e.
\begin{equation}
\{a\cdots(\ell)\cdots a\}=a \cdots (\ell)\cdots a  \ .
\end{equation}

For $r>0$, inequalities can be implemented by using them as Boolean conditions in Mathematica sums of the form
\begin{equation}
\text{Sum[ Sum[}\cdots\text{Boole[\,cond1\,,\,cond2\,,...\,]}\ a{\cdots} a\cdot\delta \cdots \delta\,,\{i,1,\ell\}],],\{j,1,\ell\}]\cdots] \ .
\end{equation}

\vspace{0.1in}

To delineate the index inequalities, first, draw a $(\ell + 1) (\ell + 1)$
matrix with rows and columns of cells labeled by indices as the first row and first column, in the order of the first $(\ell - 2 r)$  $a$'s and then the $r$ deltas. (See example in Table \ref{tab-93}.)  All cells of this matrix on the diagonal and below
will not be used.
Fill the one-up diagonal of the
upper-left rectangle of size $(\ell-2r)\times (\ell-2r)$ with less-than signs.  In the top right part of the matrix, when
$(\ell-2r)\ne 0$,
there will be a block of cells forming a rectangle with $(\ell-2r)\times 2r$ cells. Fill this block of cells with ``not equal'' signs ($\neq$), indicating
that the corresponding row indices cannot be equal to the column index. In the upper triangle of the matrix, for $r>0$, there will be square blocks of 2x2 cells each corresponding to the row and column indices for the deltas ($\delta$). Fill these 2x2 blocks with ``not equal'' signs.
Significant speed is gained
by limiting the range of the indices. Time is also reduced by taking
advantage of the fact that Boole operations
tests just up to the first false.

\begin{center}
\begin{table}
\caption{Table of index inequalities for $\ell=9,\ r=3$:}
 \begin{tabular}{|p{1.2cm}||p{0.6cm}|p{0.6cm}|p{0.6cm}||p{0.6cm}|
   p{0.6cm}||p{0.6cm}|p{0.6cm}||p{0.6cm}|p{0.6cm}|}
  \hline
  \bf{} & \bf{\ i}& \bf{\ j}& \bf{\ k}& \bf{\ l}& \bf{\ m}&
  \bf{\ n} & \bf{\ o} & \bf{\ p} & \bf{\ q}\\
  \bf{} & \bf{\ $i_1$}& \bf{\ $i_2$}& \bf{\ $i_3$}& \bf{\ $j_1$}& \bf{\ $k_1$}&
  \bf{\ $j_2$} & \bf{\ $k_2$} & \bf{\ $j_3$} & \bf{\ $k_3$}\\
  \hline\hline
  \textbf{\ i \ \ \ $i_1$}  & &\ \ $<$& &\ \ =\hspace{-0.1in}/&
  \ \ =\hspace{-0.1in}/&\ \ =\hspace{-0.1in}/&\ \ =\hspace{-0.1in}/&\ \ =\hspace{-0.1in}/&\ \ =\hspace{-0.1in}/\\
  \hline
  \textbf{\ j \ \ \ $i_2$} &  &  & \bf{\ \ $<$} &  \ \ =\hspace{-0.1in}/ & \ \ =\hspace{-0.1in}/&\ \ =\hspace{-0.1in}/&\ \ =\hspace{-0.1in}/&\ \ =\hspace{-0.1in}/&\ \ =\hspace{-0.1in}/\\
  \hline
  \textbf{\ k \ \ \,$i_3$} &  &  &  & \ \ =\hspace{-0.1in}/ & \ \ =\hspace{-0.1in}/
  &\ \ =\hspace{-0.1in}/&\ \ =\hspace{-0.1in}/&\ \ =\hspace{-0.1in}/&\ \ =\hspace{-0.1in}/\\
  \hline\hline
  \textbf{\ l \ \ \ $j_1$} &  &  &  &  & \ \ $<$&\ \ =\hspace{-0.1in}/&\ \ =\hspace{-0.1in}/&\ \ =\hspace{-0.1in}/&\ \ =\hspace{-0.1in}/\\
  \hline
  \textbf{\,m \ \,\,$k_1$} &  &  &  &  & &\ \ =\hspace{-0.1in}/&\ \ =\hspace{-0.1in}/&\ \ =\hspace{-0.1in}/&\ \ =\hspace{-0.1in}/\\
  \hline\hline
  \textbf{\ n \ \ \,$j_2$} & & & & & & &\ \ $<$&\ \ =\hspace{-0.1in}/&\ \ =\hspace{-0.1in}/\\
  \hline
  \textbf{\ o \ \ \,$k_2$} &  &  &  &  &
  & & &\ \ =\hspace{-0.1in}/&\ \ =\hspace{-0.1in}/\\
  \hline\hline
  \textbf{\ p \ \ \,$j_3$} &  &  &  &  & & & & &\ \ $<$\\
  \hline
  \textbf{\ q \ \ \,$k_3$} &  &  &  &  & & & & & \\
  \hline
 \end{tabular}
 \label{tab-93}
 \end{table}
 \end{center}
Below is an example Mathematica code to find all terms in
$\{a\cdots(\ell-2r)\cdots a\ \delta\cdots (r)\cdots \delta\}$ with $\ell=9$, $r=3$:

\vspace{0.05in}

\begin{center}
 \begin{verbatim}
              p93 = FortranForm[Sum[Sum[Sum[Sum[Sum[Sum[Sum[Sum[Sum[
                    Boole[i != l && i != m && i != n && i != o && i != p &&
                          i != q && j != l && j != m && j != n && j != o &&
                          j != p && j != q && k != l && k != m && k != n &&
                          k != o && k != p && k != q && l != n && l != o &&
                          l != p && l != q && m != n && m != o && m != p &&
                          m != q && n != p && n != q && o != p && o != q]
                     a[i] *a[j] *a[k] *d[l,m] *d[n,o] *d[p,q],
                         {q, p + 1, 9}], {p, n + 1, 8}], {o, n + 1, 7}],
                         {n, l + 1, 6}], {m, l + 1, 5}], {l, 1, 4}],
                         {k, j + 1, 9}], {j, i + 1, 8}], {i, 1, 7}]];
 \end{verbatim}
\end{center}

The output from the Mathematica code for p93 will have 1260 terms, which Mathematica took two and a half seconds to find on a 386-i7-PC. The compiled assembly language program took less than 25 milliseconds for the same calculation.  This is a hundred fold
speed advantage. Of course, a compiled program in C or C++ should be
faster than Mathematica, but with human intelligence in coding, never faster
than compiled Assembly.

\section{conclusions}

Writing an involved algebraic assembly code, or, for that matter,
composing any long stretch of assembly code, is challenging. However, for problems involving
perhaps hundreds of thousands of actions
on symbols, coding which directly manipulates objects
at the machine instruction level can make executables which run
significantly faster than compiled higher-level code simply
because, so far, humans are better at seeing opportunities
for optimization of a code for a particular CPU than the current
optimization strategies used in the best compilers for higher-level languages.

\appendix

\section{Assembly Code for Cartesian Harmonic Tensors}
\label{app-ass-code}

This code is available in the ancillary material associated
with this paper.

\bibliographystyle{apalike}



\end{document}